\input{aipcheck}

\documentclass[final]{aipproc}

\usepackage{slashed}

\newcommand{\pslash}{\not{\hbox{\kern-2.pt $p$}}}
\newcommand{\kslash}{\not{\hbox{\kern-1.5pt $k$}}}
\newcommand{\qslash}{\not{\hbox{\kern-1.5pt $q$}}}
\newcommand{\lslash}{\not{\hbox{\kern-.3pt $l$}}}
\newcommand{\llslash}{\not{\hbox{\kern-.3pt $l_{1}$}}}
\newcommand{\lllslash}{\not{\hbox{\kern-.3pt $l_{2}$}}}

\layoutstyle{6x9}

\begin{document}

\title{Higgs boson at LHC: a diffractive opportunity}


\author{M.B. Gay Ducati and G.G. Silveira}{
address={High Energy Physics Phenomenology Group, UFRGS, \\ Caixa Postal 15051, CEP 91501-970 - Porto Alegre, RS, Brazil.}
}

\begin{abstract}
An alternative process is presented for diffractive Higgs boson production in peripheral $pp$ collisions, where the particles interact through the Double Pomeron Exchange. The event rate is computed as a central-rapidity distribution for Tevatron and LHC energies leading to a result around 0.6 pb, higher than the predictions from previous approaches. Therefore, this result arises as an enhanced signal for the detection of the Higgs boson in hadron colliders. The predictions for the Higgs boson photoproduction are compared to the ones obtained from a similar approach proposed by the Durham group, enabling an analysis of the future developments of its application to $pp$ and $AA$ collisions.
\end{abstract}

\maketitle

\section{INTRODUCTION}

A new way to produce the Higgs boson in Peripheral Collisions is calculated assuming that, at high energies, the protons interact through the Double Pomeron Exchange (DPE) \cite{KMR-1997}. The interaction will occur between the colliding proton and the emitted photon from the electromagnetic field around the second proton \cite{hencken}. Thus, the only way to have an interaction by DPE in a photon-proton process is to consider the photon splitting into a quark-antiquark pair, which enables the use of the impact factor formalism. Adopting this mechanism for an elastic process, the final state of the exclusive event will be characterized by the presence of rapidity gaps between the Higgs and the particle under interaction, as shown in Fig.~{\ref{fig:foto-part}}a. Therefore, the goal is to compute the event rate for the central Higgs production in the $\gamma p$ process.

\section{THE PARTONIC PROCESS}

\begin{figure}[t]
\centering
\resizebox{\textwidth}{!}{
\scalebox{00.30}{\includegraphics[bb = 132 450 458 750 , scale=00.55]{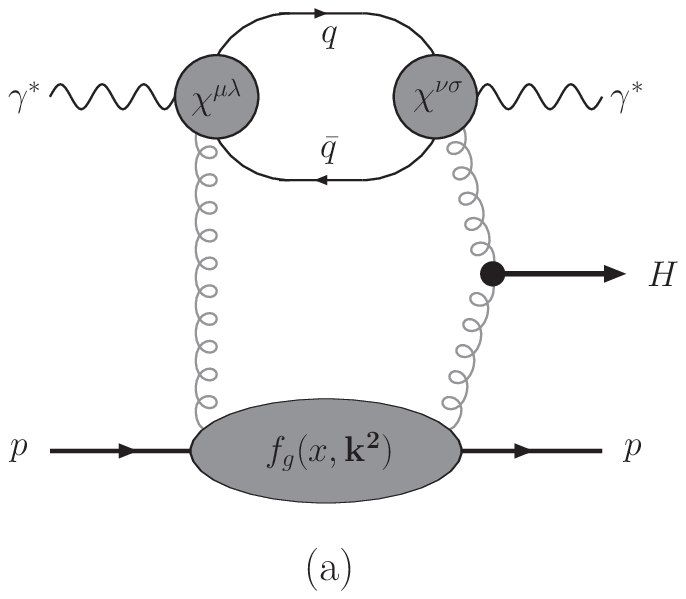}}
\scalebox{00.30}{\includegraphics[bb = 72 145 494 453 , scale=00.55]{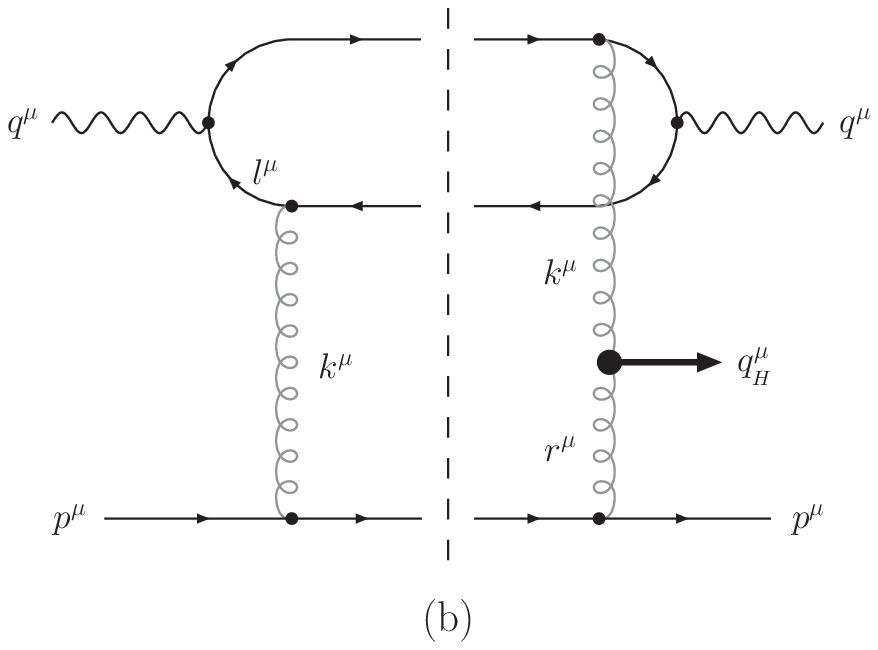}}
}
\caption{\label{fig:foto-part} The figure in the left shows the diagram representing the $\gamma^{*} p$ subprocess. The upper blobs represent the photon-gluon coupling, and the lower one the gluon distribution function in the proton. In the right, the Feynman diagram represents the diffractive Higgs boson photoproduction.}
\end{figure}

The event rate for the Higgs boson production is computed perturbatively at partonic level, where the process $ \gamma^{*} q \to \gamma^{*} + H + q$, shown in Fig.~\ref{fig:foto-part}b, has a leading contribution. In order to use the Cutkosky rules, a central line cuts the diagram in two distinct parts, and the imaginary part of the amplitude is given by
\begin{eqnarray}
{\rm{Im}}A = \frac{1}{2} \int d(PS)_{3} \, {\cal{A}}_{L} \, {\cal{A}}_{R},
\label{cut-rule}
\end{eqnarray}
with ${\cal{A}}_{L}$ and ${\cal{A}}_{R}$ being the amplitudes in the left- and the right-hand sides of the cut, respectively, and $d(PS)_{3}$ is the volume element of the three-body phase space. Moreover, the fermion loop is cut in two parts, each one representing the splitting of the photon.

Performing the integration of Eq.(\ref{cut-rule}), one gets the following imaginary part of the scattering amplitude in the transversal mode of the incoming photon \cite{nos}
\begin{eqnarray}
\mbox{Im}\,A_{T} \propto - \frac{M^{2}_{H}}{\pi v} \frac{C_{F}}{N_{c}} \, s \int  \frac{d\mathbf{k}^{2}}{\mathbf{k}^{6}} \! \left[ 1 + \frac{24\mathbf{k}^{8}  -  226Q^{2}\mathbf{k}^{6}  -   733Q^{4}\mathbf{k}^{4}  -  670Q^{6}\mathbf{k}^{2}  -  186Q^{8}}{24Q^{8}  + 72Q^{6}\mathbf{k}^{2}  + 72Q^{4}\mathbf{k}^{4}  + 24Q^{2}\mathbf{k}^{6}} \, \right] \!\! . \!\!
\label{amp-im-3}
\end{eqnarray}
The production vertex $gg \to H$ is approximated for the production of a not too heavy Higgs boson (M$_{H}$ $\leq$ 200 GeV) \cite{forshaw-KMR}.

The main feature obtained in this result is the sixth-order $\mathbf{k}$-dependence, distinct from the result of the Durham group, which presents a fourth-order dependence. This difference appears due to the presence of the photon in the process, which simplifies the result by the existence of only one gluon distribution function in the process, although introducing a more complicated expression with a $Q^{2}$-dependence.

\section{PHOTON-PROTON COLLISIONS}

Introducing the Higgs boson production in Peripheral Collision \cite{hencken}, only the electromagnetic interaction occurs between the ions under collision, with an impact parameter $ |\vec{b}| \sim 2R$.  In $pp$ collisions, the range of the photon virtuality is determined by the upper value $Q^{2} < R_{p}^{-2} \sim$ 10$^{-2}$ GeV$^{2}$.

For a realistic case of $\gamma p$ interaction in Peripheral Collisions, one replaces the contribution of the $qg$ vertices by a partonic distribution in the proton \cite{KMR-CanThe}.

Nevertheless, to take into account the gluon ladder coupled to the proton, one needs to assume a small momentum fraction, like $x \sim 0.01$, such that one safely identifies the distribution function $f_{g}(x,\mathbf{k}^{2})$ \cite{KMR-1997}. Finally, the event rate reads
\begin{eqnarray}
\left. \frac{d\sigma}{dy_{H}dt} \right|_{t,y_{_{H}}  = 0}  \propto  \frac{S^{2}_{gap}}{b }  \left[ \int_{\mathbf{k}^{2}_{0}}^{\infty}  \frac{d\mathbf{k}^{2}}{\mathbf{k}^{6}} \; f_{g}(x,\mathbf{k}^{2}) \; X(\mathbf{k}^{2},Q^{2}) e^{-S(\mathbf{k}^{2},M^{2}_{H})}  \right]^{2} ,
\label{final-eq1}
\end{eqnarray}
where $X(\mathbf{k}^{2},Q^{2})$ is the function inside the brackets in Eq.(\ref{amp-im-3}), and a cutoff was included in the integration over the gluon momentum to avoid infrared divergences \cite{KMR-CanThe}. Furthermore, many gluon emissions from the production vertex are suppressed by the inclusion of the factor $e^{-S}$, and the Rapidity Gap Survival Probability (GSP) is taken into account to correctly predict the cross section. For the diffractive Higgs production, the GSP is 3\% for LHC ($\sqrt{s} = 14\textrm{ TeV}$) and 5\% for Tevatron ($\sqrt{s} = 1.96\textrm{ TeV}$) \cite{KMR,gotsman}.

\section{NUMERICAL RESULTS}

\begin{figure}[t]
\centering
\resizebox{\textwidth}{!}{
\rotatebox{-90}{\scalebox{00.31}{\includegraphics*[86pt,25pt][568pt,725pt]{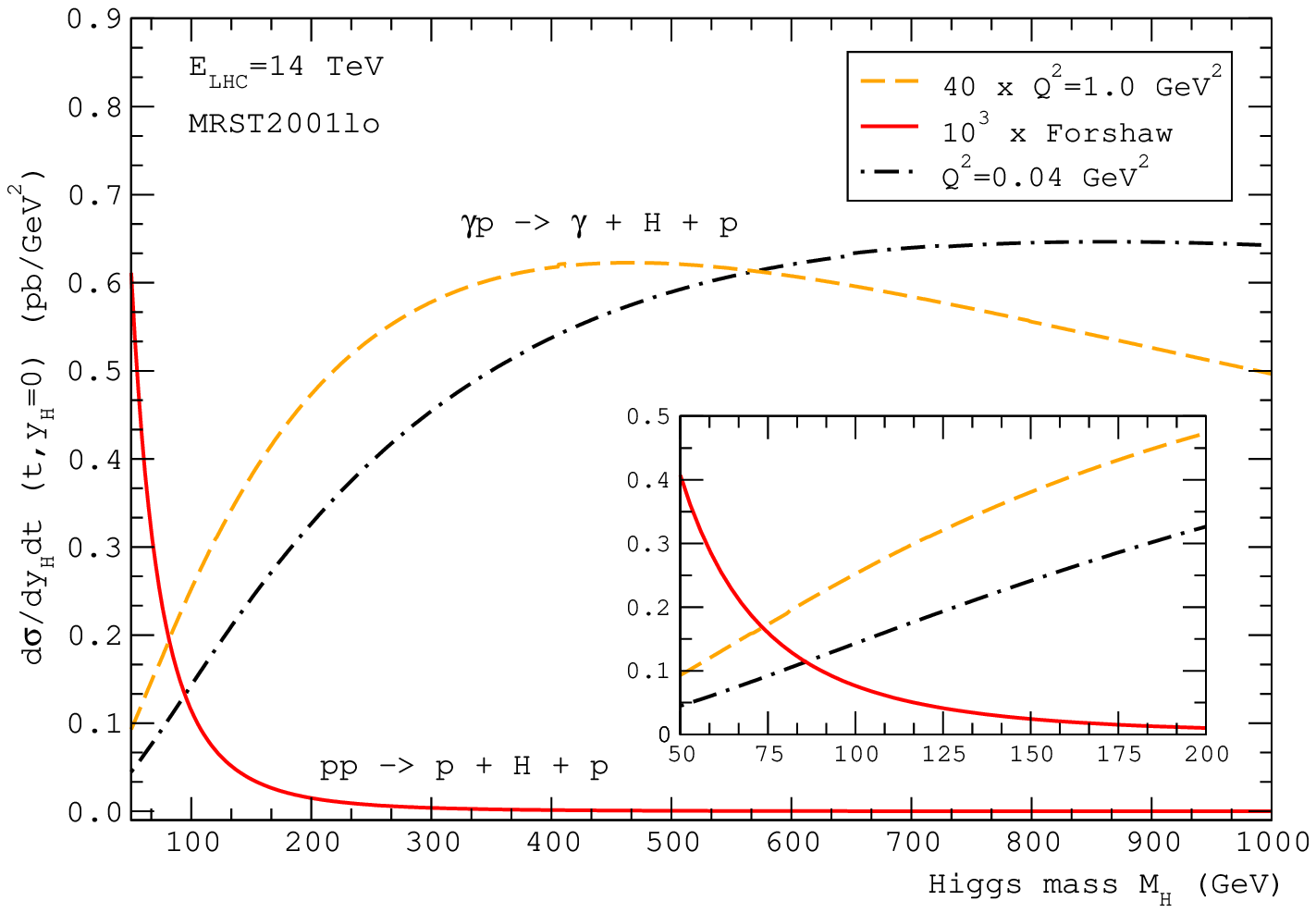}}}
\hfill
\rotatebox{-90}{\scalebox{00.30}{\includegraphics*[81pt,23pt][575pt,717pt]{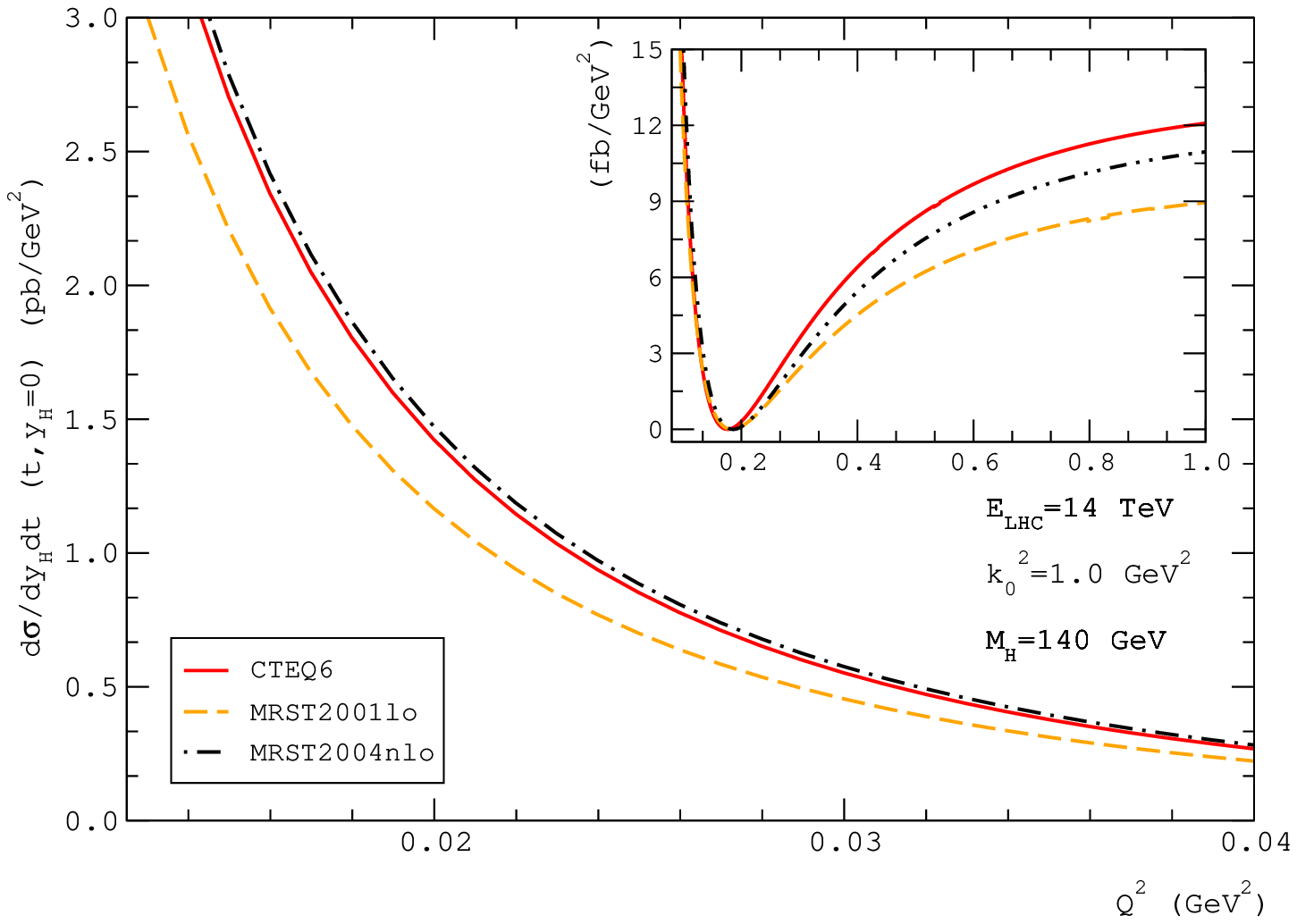}}}
}
\caption{\label{fig:forshaw} Event rate $ d\sigma/dy_{H}$ ($y_{H}$=0) for LHC energy in function of $M_{H}$ in two distinct ranges. The results are obtained using the MRST2001 parametrization in LO approximation. The results are compared with previous predictions carried out in Ref.\cite{forshaw-KMR}. Moreover, the dependence of the event rate on the photon virtuality for different parametrizations is presented.}
\end{figure}

The differential cross section in central-rapidity for LHC is computed using the parametrization MRST2001 in leading-order approximation for the gluon distribution function, taking an initial momentum of $\mathbf{k}^{2}_{0} = 1 \textrm{ GeV}^{2}$. The results are shown in Fig.~\ref{fig:forshaw} in function of the Higgs boson mass and compared to previous results \cite{forshaw-KMR}. As the mass increases the curve seems to diverge, however for heavier masses  the growing is reduced due to the Sudakov form factors. The behavior of the photoproduction results is expected not to fit like the results of direct $pp$ collisions, since the photoproduction approach assumes only one PDF to compute the event rate. Analyzing the dependence on the photon virtuality, the right graph of Fig.~\ref{fig:forshaw} shows a fast decreasing of the event rate to zero and a subsequent growing with distinct rates for each parametrization.

\begin{figure}[t]
\centering
\resizebox{\textwidth}{!}{
\rotatebox{-90}{\scalebox{00.33}{\includegraphics*[80pt,25pt][555pt,723pt]{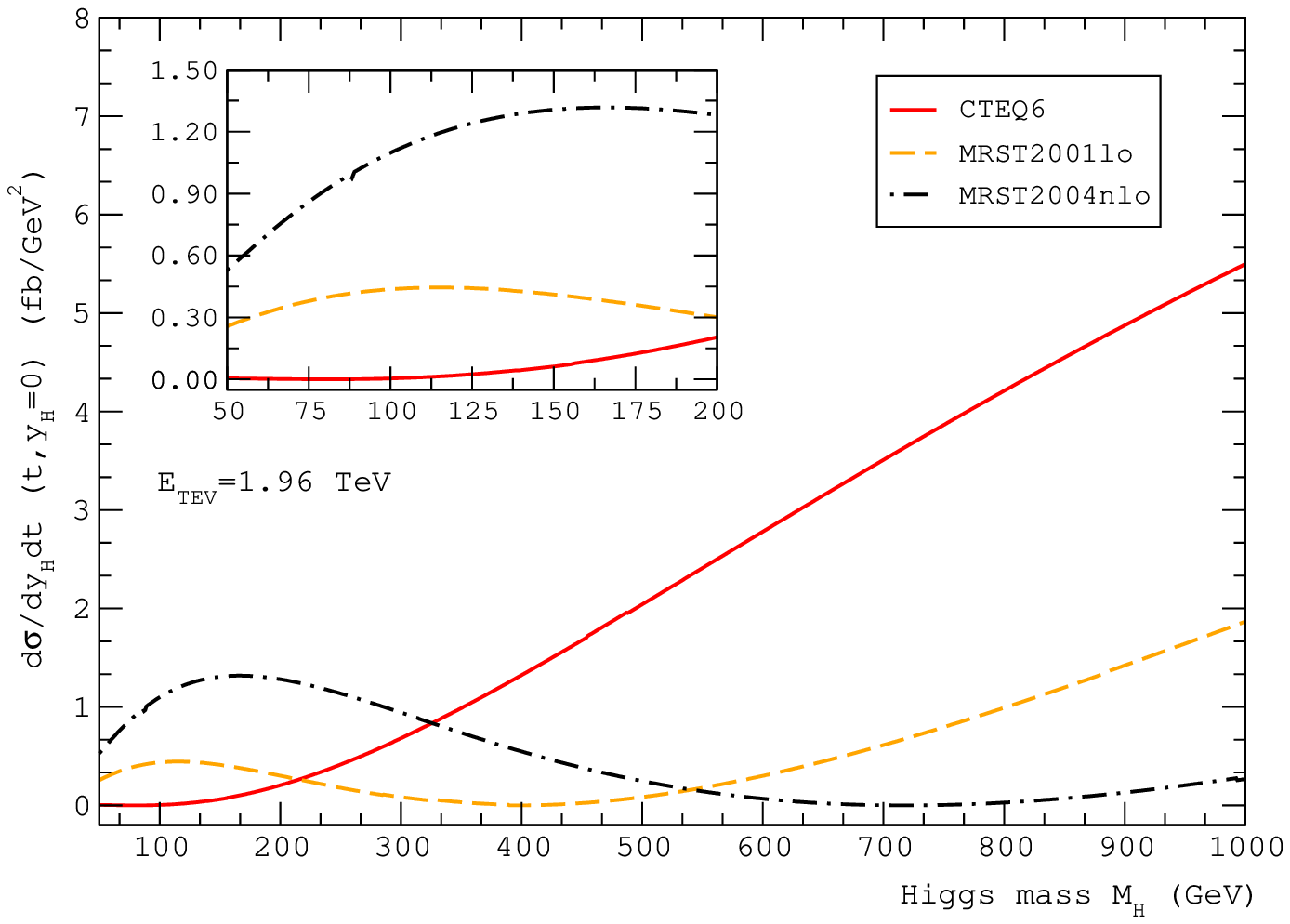}}}
\hfill
\rotatebox{-90}{\scalebox{00.33}{\includegraphics*[80pt,25pt][550pt,723pt]{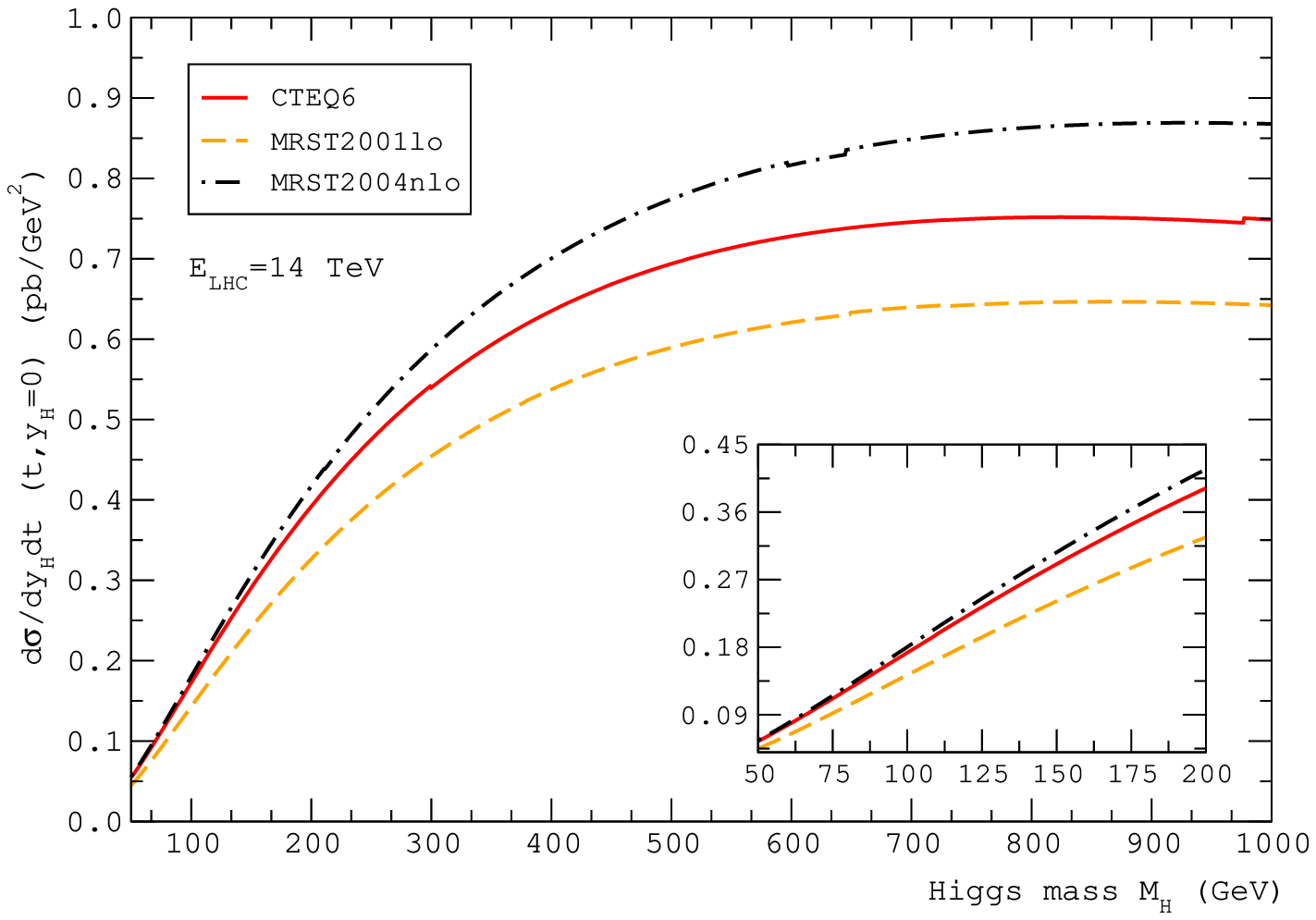}}}
}
\caption{\label{fig:fig5} Event rate in function of the $M_{H}$ for Tevatron and LHC, taking the results in an intermediary mass range and an extended one, describing the effect of distinct parametrizations of the PDF's.}
\end{figure}

Moreover, the event rate is calculated using an LO distribution and two distinct NLO distributions (MRST2004 and CTEQ6) and all distributions were evolved from an initial momentum $\mathbf{k}_{0}^{2} = 1$ GeV$^{2}$. The results are shown in Fig.~\ref{fig:fig5} for two different mass ranges and taking predictions for Tevatron and LHC energies. As one can see, the results in Tevatron have a non-uniform behavior since the necessary value for the momentum fraction is not reached (at most $x \approx 0.05$). Conversely, $x$ is easily reached in LHC and the results show a uniform growth with $M_{H}$ as well as a difference in the results among the NLO and LO distributions as the mass increases. Inspecting the sensitivity of the results on the cutoff $\mathbf{k}^{2}_{0}$, the results for LHC with $\mathbf{k}^{2}_{0} = 1$ GeV$^{2}$ are two times higher than the ones for $\mathbf{k}^{2}_{0} = 2$ GeV$^{2}$. Therefore, it shows that these results are three times less sensitive than the KMR results.

\section{CONCLUSIONS}

A new way to produce the Higgs boson was explored in Peripheral Collisions with the proposal of calculate perturbatively the event rate for diffractive production by DPE. The numerical results obtained from this approach estimate some predictions for the Higgs production at LHC with a reasonable event rate compared to previous results. The event rate was obtained in the $\gamma p$ interaction with a dependence on $\mathbf{k}^{-6}$, unlike the result obtained for direct $pp$ collisions \cite{forshaw-KMR,levin}. To effectively compare the results presented here with those of the Durham group, a distribution function for the photons into the proton should be introduced and then to compute the results for peripheral proton-proton collisions. Therefore, the results show the possibility of Higgs production through Peripheral Collisions at LHC with a signal strong enough to be detected.

\begin{theacknowledgments}
This work was partially supported by CNPq (GGS and MBGD).
\end{theacknowledgments}

\end{document}